\documentclass[usenatbib,useAMS,referee]{mn2e}
\usepackage{amssymb}
\usepackage{graphicx}
\usepackage{psfrag}

\newcommand\umin{{\,u_{\rm min}}}
\newcommand\mas{{\,\rm mas} }
\newcommand\kpc{{\,\rm kpc} }
\newcommand\Dd{D_{\rm d}}

\newcommand\Ds{D_{\rm s}}
\newcommand\zs{{z_{\rm s}}}
\newcommand\zc{{z_{\rm s, 0}}}
\newcommand\rE{r_{\rm E}}
\newcommand\rs{\rho_{\star}}
\newcommand\calK{{{\cal K}_{\rm cr}}}
\newcommand\calKx{{{\cal K}_{x}}}
\newcommand\calKy{{{\cal K}_{y}}}
\newcommand\thetaE{\theta_{\rm E}}
\newcommand\thetas{\theta_{\rm s}}
\newcommand{\beq}{\begin{equation}}              
\newcommand{\beqa}{\begin{eqnarray}}             
\newcommand{\eeq}{\end{equation}}                
\newcommand{\eeqa}{\end{eqnarray}}               
\newcommand{\eeqi}[1]{\quad#1\end{equation}}     
\newcommand{\eeqai}[1]{\quad#1\end{eqnarray}}    

\newcommand{\ii}{\mbox{i}}
\newcommand{\dd}{\mbox{d}}
\newcommand{\sii}{\mbox{\scriptsize i}}

\title
[Interferometric Visibility and Closure Phase of Microlensing Events with
  Finite Source Size]
{Interferometric Visibility and Closure Phase of Microlensing Events with
  Finite Source Size}
\author[Rattenbury \& Mao]
{Nicholas J. Rattenbury and Shude Mao
\thanks{(njr, smao)@jb.man.ac.uk} \\
{University of Manchester, Jodrell Bank Observatory,
  Macclesfield, Cheshire SK11 9DL, UK}
}

\date{
Accepted ........
Received .......;
in original form ......}

\pubyear{2005}

\begin{document}
\maketitle

\begin{abstract}
Interferometers from the ground and space will be able to
resolve the two images in a microlensing event. This will at least partially
lift the inherent degeneracy between physical parameters in microlensing events. To increase the
signal-to-noise ratio, intrinsically bright events with large
magnifications will be preferentially selected as targets. These events
may be influenced by finite source size effects both photometrically
and astrometrically. 
Using observed finite source size events as examples, 
we show that the fringe visibility can be affected by $\sim$ 5-10\%, and the
closure phase by a few degrees: readily detectable by ground and space interferometers. Such detections
will offer unique information about the lens-source trajectory relative
to the baseline of the interferometers. Combined with 
photometric finite source size effects, interferometry offers a way to measure the 
angular sizes of the source and the Einstein radius accurately.
Limb-darkening changes the visibility by a small amount compared with
a source with uniform surface brightness, marginally detectable with ground-based instruments.
We discuss the implications of our results for the plans
to make interferometric observations of future microlensing events.
\end{abstract}

\maketitle

\section{Introduction}

The most serious problem in gravitational microlensing is the following well-known
degeneracy:  the only physical parameter that
can be extracted from an observed light curve is the Einstein radius crossing
time, which depends on the lens mass, the distances to the lens and
source, and the transverse velocity. This implies the lens mass cannot be
uniquely determined from an observed light curve (see \citealt{Pac96}
for a review). Information about the lens population has to be decoded
statistically. There are various ways to break the degeneracy:
astrometric signatures of microlensing events offer an exciting
possibility to do this (\citealt{Gou92, Hos93, Hog95, Miy95, Wal95, mir96, Pac98}). To completely
break the degeneracy, one must measure both the lens-source relative
parallax, $\pi_{\rm E}$, and the angular size of the Einstein radius, $\thetaE$.

Future space interferometers like the Space Interferometry Mission 
(SIM, \citealt{Sha04}) and \emph{Gaia} (\citealt{Per05}) will offer astrometric accuracies of a few
microarcseconds and therefore would be ideal for astrometric microlensing.
Unfortunately these two satellites will not be launched
until the beginning of the next decade. Note that while these two space interferometers will be sensitive to the movement of the light centroid during a microlensing event, they will not be able to resolve the two microimages. Rapid progress also has been
made on the ground, in particular with the Keck interferometer (\citealt{Boo99}), 
the Very Large Telescope Interferometer (VLTI, \citealt{Mar98}) and the Center for High Angular Resolution Astronomy (CHARA) array (\citealt{Bru05}).
For massive lenses such as stellar mass black holes, the two
micro-images, separated by $\sim$ one  angular
Einstein radius ($\sim$ a few milliarcseconds), becomes comparable to
the resolution of VLTI, $\lambda/B \approx 5\mas$, for
$\lambda=2.2\,\mu$m ($K$-band), and baseline $B=100$\,m. 
\citet{Del01} first pointed out that for such lenses, 
the fringe visibility decreases as the two microimages become resolved
by the interferometer. The change in the fringe visibility therefore
offers a useful way to determine the angular Einstein radius $\thetaE$.
\citet{Dal02} studied the (closure) phase  as an alternative way of
determining $\thetaE$; they also carefully considered the feasibility
of observing interferometric signals from the ground.

In fact, ground-based interferometric observations have already been attempted
with the VLTI, albeit with an instrument still under commissioning, 
for the bright event OGLE-2005-BLG-099 on June 21/22 2005
(A. Richichi 2005, private communication). The
event reached a $K$-band magnitude of 7-8 at the peak of the light curve.
Although this effort was unsuccessful, it is likely 
that the interferometric signatures of future
microlensing events will be observed. The requirement to obtain sufficient
photon statistics implies that the microlensing events picked for interferometric
observations are likely to be intrinsically bright, and with high
magnifications which will further boost the signal-to-noise ratio. Such events are most
likely to exhibit finite source size effects. The photometric finite
source size effects have been predicted (\citealt{Gou94, WM94, Nem94})
and observed for several single microlensing events (e.g, MACHO 95-BLG-30,
\citealt{Alc97}; OGLE-2003-BLG-262, \citealt{Yoo04}; OGLE-2003-BLG-238,
\citealt{Jia04}; see Table \ref{table:single} for details)
and a number of binary events (e.g.,
97-BLG-28, \citealt{Alb99}; OGLE-1999-BUL-23, \citealt{Alb01}; 
EROS-BLG-2000-5, \citealt{An02}, \citealt{Cas01}; MOA-2002-BLG-33,
\citealt{Abe03}, \citealt{Rat05}). The finite source
size effects have been used to constrain the limb-darkening profile 
(e.g., \citealt{Alb01}, \citealt{Abe03}) and geometric shape of the
lensed stars (\citealt{Rat05}), and in one case,
combined with other exotic effects, yielded the first unique lens mass
determination (\citealt{An02}).

Astrometrically, the motion of the light centroid is also affected by
finite source size effects (\citealt{MW98}) and can be used to determine accurately the source
size. This paper explores how the interferometric signals are
affected by the finite source size in single microlensing events.
We show that signatures in the fringe visibility  and closure phase
can in principle provide an independent way to measure the source size,
complementing the determination from the photometric light curve.
The structure of the paper is as follows. In Section~2, we introduce 
the basics of microlensing and interferometric signals. In Section~3, we present our
main results. And finally in Section~4, we summarise our results
and discuss the implications of our results for the plans
to make interferometric observations of future microlensing events.

\section{Basics of Microlensing and Interferometry}

\subsection{Lens Equation}

For convenience, we normalise all the lengths to the Einstein radius in
the lens plane, $\rE$, 
and all the angles to the angular Einstein radius, $\thetaE$. They are respectively
given by
\beq
\rE = \left({4GM \over c^2} {\Dd (\Ds-\Dd) \over \Ds}\right)^{1/2}
= 4\, {\rm AU} \left({M\over M_\odot}\right)^{1/2} \left({\Ds \over
  8\kpc}\right)^{1/2} \left( {x(1-x) \over 1/4}\right)^{1/2},
\eeq
and
\beq
\thetaE=\left({4GM \over c^2} {\Ds-\Dd \over \Dd \Ds}\right)^{1/2}
= 1 \mas \left({M\over M_\odot}\right)^{1/2} \left({\Ds \over
  8\kpc}\right)^{1/2} \left( {1-x \over x}\right)^{1/2},
\eeq
where $M$ is the lens mass, $\Dd$ and $\Ds$ are the distances to the
lens and source respectively, and $x=\Dd/\Ds$.

With these units, the lens equation in complex notation can be simply written as 
(\citealt{Wit90})
\beq
\zs = z - {1 \over \bar{z}},
\eeq
where $\zs$ is the source position, $z$ is the image
position, $\bar{z}$ is the complex conjugate of $z$, and the lens is at the origin.
The lens equation always has two solutions (images); their
positions and absolute magnifications are given by (\citealt{Lie64})
\beq
z_{+,-} = {\zs \over 2} \left[ 1 \pm \sqrt{1 + {4 \over |\zs|^2}} \,\, \right], ~~
\mu_{+, -} = \pm{1\over 2} + {|\zs|^2+2 \over 2 |\zs| \sqrt{4 + |\zs|^2}},
\label{eq:images}
\eeq
where the $+$ and $-$ signs correspond to the positive and negative parity
images respectively. 

Owing to the relative motions of the lens, source and the observer, the
magnification changes as a function of time. The resulting light curve
is usually symmetric, achromatic and due to the low microlensing
probability (approximately one in a million), non-repeating. For most
microlensing events, the light curves are well-approximated using point
sources, but for a small fraction of events, particularly for
intrinsically bright evens with high magnification, 
the finite source size effects become important, including for
interferometric signals, the subject of this paper.

\begin{table}
\begin{tabular}{ccccccccc}
\hline
Event & $t_E ({\rm d})$ & $\thetaE \,(\mas)$ & $u_{\rm min}$ &$I_{\rm
  min}$ & $\rs$ & $u_1$ & $u_2$ & References\\
\hline
MACHO-1995-BLG-30 & $67.28\pm 0.27$ & $\thetaE=0.420^{+0.68}_{-0.32}$
& 0.05408& $11.5^{\dagger}$ & 0.075 & 0.72$^{\star}$& 0.05$^{\star}$& \citet{Alc97}\\
OGLE-2003-BLG-238 & $38 \pm 0.2$ & $\thetaE=0.650\pm 0.056$ & 0.002 &
10.3 & 0.01282 & 0.5807 &0.0 & \citet{Jia04} \\
OGLE-2003-BLG-262 & $12.5 \pm 0.1$ & $\thetaE=0.195\pm 0.017$ & 0.0365 & 10.4
& 0.0605 & 0.7027 &0.0 & \citet{Yoo04} \\
\hline
\end{tabular}
\caption{Microlensing parameters for finite source size, single lens
 events. $t_E$ is the Einstein radius crossing time, $\thetaE$ is the
angular Einstein radius, $u_{\rm min}$ is the minimum impact parameter,
and $I_{\rm min}$ is the $I$-band magnitude at the peak of the
light curve. $\rs$ is the source radius, and $u_1$ and $u_2$
limb-darkening parameters defined in eq (\ref{eq:limb}).
$^{\dagger}$Assumes $V-I = +3.39$ for stellar type M4~III.
$^{\star}$Values for the MACHO R-band.
}
\label{table:single}
\end{table}

\subsection{Interferometric fringe visibility}

For a source with an intensity distribution $I(\vec{z})$ 
(hereafter, we will use the complex notation, e.g. $z$, and the vector notation,
$\vec{z}$, interchangeably), the fringe
pattern is just the Fourier transform of the source brightness profile divided
by the total intensity
\beq
\hat{V}={\int I(\vec{z}) e^{-{2 \pi i \over \lambda} \vec{z}\cdot \vec{B}} ~d^2  \vec{z}
\over  \int I(\vec{z})~ d^2\vec{z}},
\label{eq:Vdef}
\eeq
where $\lambda$ is the wavelength, $\vec{B}$ is the baseline vector
of the interferometer, and $\vec{z}$ is the vector to an infinitesimal
element of the source.

In microlensing, if the two images are unresolved (point-like) by the
interferometer, then the complex visibility is simply given by
\beq
\hat{V}={\mu_+ e^{-{2 \pi i \over \lambda}\thetaE \vec{z}_+\cdot\vec{B}}
+\mu_- e^{-{2 \pi i \over \lambda} \thetaE \vec{z}_-\cdot\vec{B}} 
\over \mu_+ + \mu_-}.
\eeq
The amplitude of the complex visibility is given by
\beq
|\hat{V}|^2 = {1 \over (R+1)^2}\left[ {1+R^2 + 2 R \cos({
\vec{\calK} \cdot \Delta \vec{z}})} \right],  ~~
\eeq
where $R=\mu_+/\mu_-$ is the magnification ratio between the two
images, $\Delta \vec{z} = {\vec{z}_+ - \vec{z}_-}$, and the dimensionless parameter 
\beq
\calK = |\vec{\cal K}_{\rm cr}|= 2 \pi {B \thetaE \over \lambda}
= 0.69 \left({B \over 100 \,{\rm m}}\right) 
\left({\thetaE \over 0.5 \,{\rm mas}}\right) \left({2.2\, \mu{\rm m}
  \over \lambda}\right),
\label{eq:calK}
\eeq
is the phase difference between two micro-images separated by one angular Einstein radius. 
We note that the visibility $|V|$ is a sensitive function of $\calK$ when $R \sim 1$.

\subsection{Three-element interferometry}
The phase measured by a single baseline interferometer suffers from the effects of atmospheric turbulence, and is consequently useless without simultaneous measurements of a phase reference source. The visibility from each telescope in a single baseline interferometer is (e.g. \citealt{Dal02}):

\beq
\overline{V} = |\hat{V}|e^{i(\phi_{12} + \phi_{1} - \phi_{2})},
\eeq
where $\phi_{12}$ is the intrinsic phase difference due to the baseline separation between elements 1 and 2, $\phi_{1}$ and  $\phi_{2}$ are random phases introduced by the time-varying refraction characteristics of the atmosphere above the two telescopes. The product of the three visibilities obtained from a three-element interferometer with the baselines arranged as a closed triangle is called the bispectrum. The phase of the bispectrum is called the closure phase and has the desirable property of being independent of the random phase differences imposed by the atmosphere above the interferometer (\citealt{Jen58}; \citealt{Cor81}). If $\phi_{ij}$ is the phase associated with the baseline between interferometer elements $i$ and $j$, and $\phi_{i}$ is the random phase due to the turbulent atmosphere above telescope $i$, then the bispectrum is:

\beq
V_{\rm bis} =   |V_{1}||V_{2}||V_{3}| e^{i(\phi_{12} + \phi_{23} + \phi_{31})}.
\eeq
The closure phase is therefore the sum of the intrinsic phases arising from the three baselines. Closure relationships were developed and are routinely used in radio interferometry (e.g. \citealt{Tho01}; \citealt{Bur02}).

\subsection{Interferometric visibility for sources with uniform surface brightness}

To see why finite source size effects may be important for
interferometric signals, let us first estimate the phase difference due to the 
finite size of images.
For very high magnification events, the two micro-images are distorted
into two long arcs. The radial width of each image is equal to the source radius (\citealt{Lie64}),
while the length of each arc is given by $\approx \mu$ times the source
radius, where $\mu=\mu_+ + \mu_-$ is the total
magnification of the two micro-images. If the interferometer is perpendicular to
the image seperation axis and we ignore the curvature of the image arcs, then the phase difference due to the length of the images is approximately
\beq
2 \pi {B \mu \thetas \over \lambda}
\approx \calK {\rs \over \umin},
\eeq
where $\rs \equiv {\thetas / \thetaE}$ is the physical source radius in units of the Einstein radius, $\mu \sim {1 / \umin}$ and $\umin \ll 1$.
Therefore, for finite source size events, where $\umin \sim \rs$, 
the phase difference parallel to the arc cannot be neglected.
On the other hand, the phase difference due to the width of the arc is 
usually smaller for the high-magnification events. 

\begin{figure}
\centering\includegraphics[width=0.8\hsize]{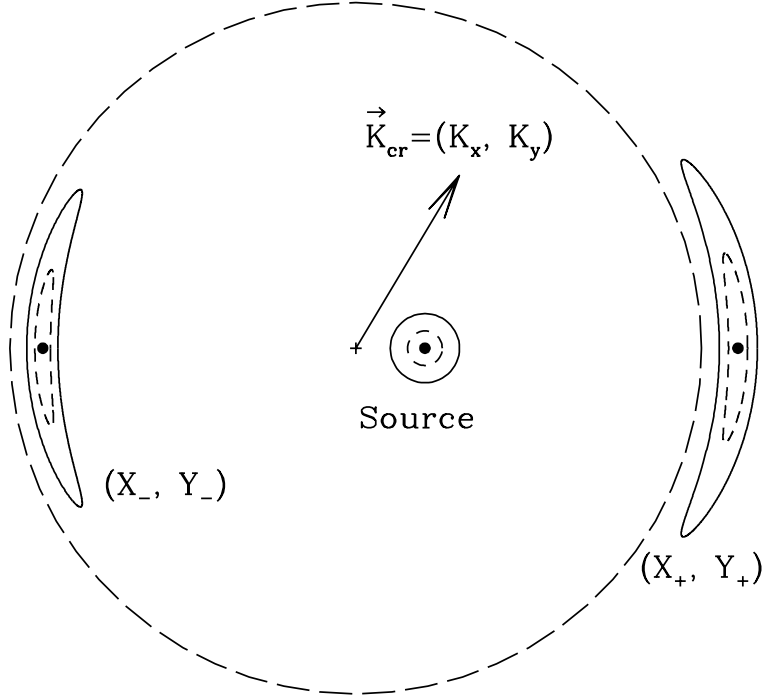}
\caption{
Illustration of the lensing geometry. The Einstein radius is shown as a long-dashed
circle with the lens at the origin, indicated by a plus symbol. The centre of the source is on the positive horizontal axis. The source
size is indicated by a small solid circle and is mapped into the two arcs
(solid curves), labelled by ($X_-, Y_-$) and ($X_+, Y_+$).
An inner dashed circle is mapped into two smaller arcs shown as dashed lines. A 
point source (indicated by a large dot) is mapped to the two large dots at the centre
of the arcs. We also indicate $\vec{\cal K}_{\rm cr}$ (defined in eq. \ref{eq:calK}) which
 is in general at some angle with respect to the $x$-axis.
}
\label{fig:geometry}
\end{figure}

Let us consider a circular source with radius $\rs$ and constant
surface brightness. The area of the source is specified by
$\zs(r, \varphi) =\zc + r e^{\sii\varphi}$, where $\zc$ is the centre of
the source, and $0 \le\varphi \le 2\pi$, and $0\le r \le \rs$. The boundary
of the source is given by $\zs(\rs, \varphi)$. For each circle of radius
$r$, from eq. (\ref{eq:images}), the parametric representation for the two
images is
\beqa
z_{+,-} (r, \varphi ) &=& \frac{\zc+ r e^{\sii\varphi}}{2} \left[
 1\pm \sqrt{1 +\frac{4}{g(\varphi)}}
~ \right] \nonumber  \\
  &=& x_{+,-} (r, \varphi ) + \ii y_{+,-} (r, \varphi ),
 \label{eq:circ}
\eeqa
where we have assumed (without losing generality) that $\zc$ is on the positive
horizontal axis, and $g(\varphi) = r^2 +\zc^2 +2 r\zc \cos
\varphi$. As we mentioned, the images generally form two arcs. The boundaries of the positive and
negative parity images are given by $r=\rs$, which we will denote as
$X_{\pm}(\varphi)$ and $Y_{\pm}(\varphi)$. The lensing geometry and notations
are illustrated in Fig.~\ref{fig:geometry}.

Since lensing conserves surface brightness, the magnification of a circular
source with uniform surface brightness is just given 
by the ratio of the area of the images to the area of the source. For
any closed curve, its area is given by an integral of the outer
boundary, hence the total magnification is given by 
%
\beq \label{mag}
\mu_{\rm tot} = \frac{1}{\pi \rs^2} \int\limits_0^{2 \pi} \left[
-Y_+(\varphi)\frac{\dd X_+(\varphi)}{\dd\varphi}
+Y_-(\varphi)\frac{\dd X_-(\varphi)}{\dd\varphi}
\right]\dd \varphi. \label{eq:mu}
\eeq
Note that the minus sign results from the fact that when $\varphi$ changes from 0 to
$2\pi$, the contour for the image with positive parity moves
counter-clockwise whereas that for the negative parity image moves clockwise.

The centroid of light can be calculated by weighting the
image positions with magnification. For a uniform surface brightness
circular source, this is given by \citet{MW98}:
\beq \label{clight}
\Delta \theta_x = \frac{1}{\pi \rs^2 \mu_{\rm tot}} \int\limits_0^{2 \pi} 
\left[
- X_+(\varphi) Y_+(\varphi)\frac{\dd X_+(\varphi)}{\dd\varphi}
+ X_-(\varphi) Y_-(\varphi)\frac{\dd X_-(\varphi)}{\dd\varphi}
\right]\dd \varphi. \label{eq:theta}
\eeq
Note that $\Delta\theta_y=0$ because of the reflection symmetry
with respect to the horizontal axis.
The integrals in eqs. (\ref{eq:mu}) and (\ref{eq:theta})  
can be evaluated analytically; the results have been
presented in \citet{WM94} and \citet{MW98}, the readers are referred to
those papers for details.

After some algebra, the complex visibility as defined in
eq. (\ref{eq:Vdef}) can be found to be 
\beq
\hat{V} = {1 \over \mu_{\rm tot} \pi \rs^2} \int\limits_0^{2 \pi} \left[
-Y_+(\varphi)\frac{\dd X_+(\varphi)}{\dd\varphi} e^{i \calKx  X_+(\varphi)}
{\sin \calKy Y_+ \over \calKy}
+Y_-(\varphi)\frac{\dd X_-(\varphi)}{\dd\varphi} e^{ i \calKx  X_-(\varphi)}
{\sin \calKy Y_- \over \calKy} 
\right]\dd \varphi,
\eeq
where $\calKx$ and $\calKy$ are the $x$ and $y$ components of $\vec{\cal  K}_{\rm cr}$.
When the source is exactly aligned with the
line of sight, one can show that the visibility can be simplified into Bessel
functions (see Appendix A). In general, the visibility appears analytically intractable.
In any case, the integral can be efficiently evaluated numerically.

\subsection{Interferometric visibility for sources with limb-darkened profiles}

For sources with limb-darkened profiles, $I(r)$, the visibility 
in general requires two-dimensional integration. The lens equation gives
the mapping for any given source position onto the lens plane, so we can
use the Jacobian to derive the area occupied by the images for 
any source element $\dd x \dd y$. Realising this,
the visibility can be cast using the Jacobian
\beq
\hat{V} = { {1 \over \mu_{\rm tot} \pi \rs^2}}
\int_0^{\rs} I(r) dr \int\limits_0^{2 \pi} \left[
-{\partial(x_+, y_+) \over \partial(r, \varphi)}  +
{\partial(x_-, y_-) \over \partial(r, \varphi)}
\right]\dd \varphi, ~~ r = \sqrt{x^2+y^2}.
\eeq
where $\mu_{\rm tot}$ is the total magnification.
The Jacobian in the above integral can be obtained through differentiation of
the image positions with respect to $r$ and $\varphi$ (in eq. \ref{eq:circ}), and the integral can be
performed numerically.

\section{Results}

We will use the first few microlensing events which
exhibit photometric extended size effects (see introduction) for illustration purposes.
We model the sources with uniform surface brightness  and limb-darkened profiles.
Following \citet{Alc97}, we model the source limb-darkening profile by
(e.g., \citealt{All73, Cla95})
\beq
\frac{I(R)}{I(0)}=
1-u_1-u_2 + u_1 \sqrt{1-R^2} + u_2 
\left(1-R^2\right). \label{eq:limb}
\eeq
where $R$ is the radial distance from the centre of the source star in
units the source physical radius, and $I(0)$ is the central surface brightness.
The parameters $u_1$, $u_2$ depend mainly on the effective temperature
and surface gravity of the source star. \citet{Jia04} and \citet{Yoo04}
restrict their limb-darkening models to those with $u_2=0$. The parameters 
for three known single events with finite source size effects are listed
in Table \ref{table:single}.

For definitiveness, we will use the VLTI as an example of a ground-based interferometer.
We adopt a baseline of 100\,m, and $\lambda=2.2\mu$m ($K$-band).
The VLTI has multiple configurations and can have two or many more baselines
(see Table 1 in \citealt{Del01} for details; see also \citealt{Dal02}).
The minimum difference in the visibility detectable by the VLTI is
estimated to be between 0.5\% and 5\% depending on the interferometer configuration, and the limit on measuring the closure phase is currently approximately $1^{\circ}$.

\subsection{Two-element interferometer}

\begin{figure}
\centering\includegraphics[width=0.49\hsize]{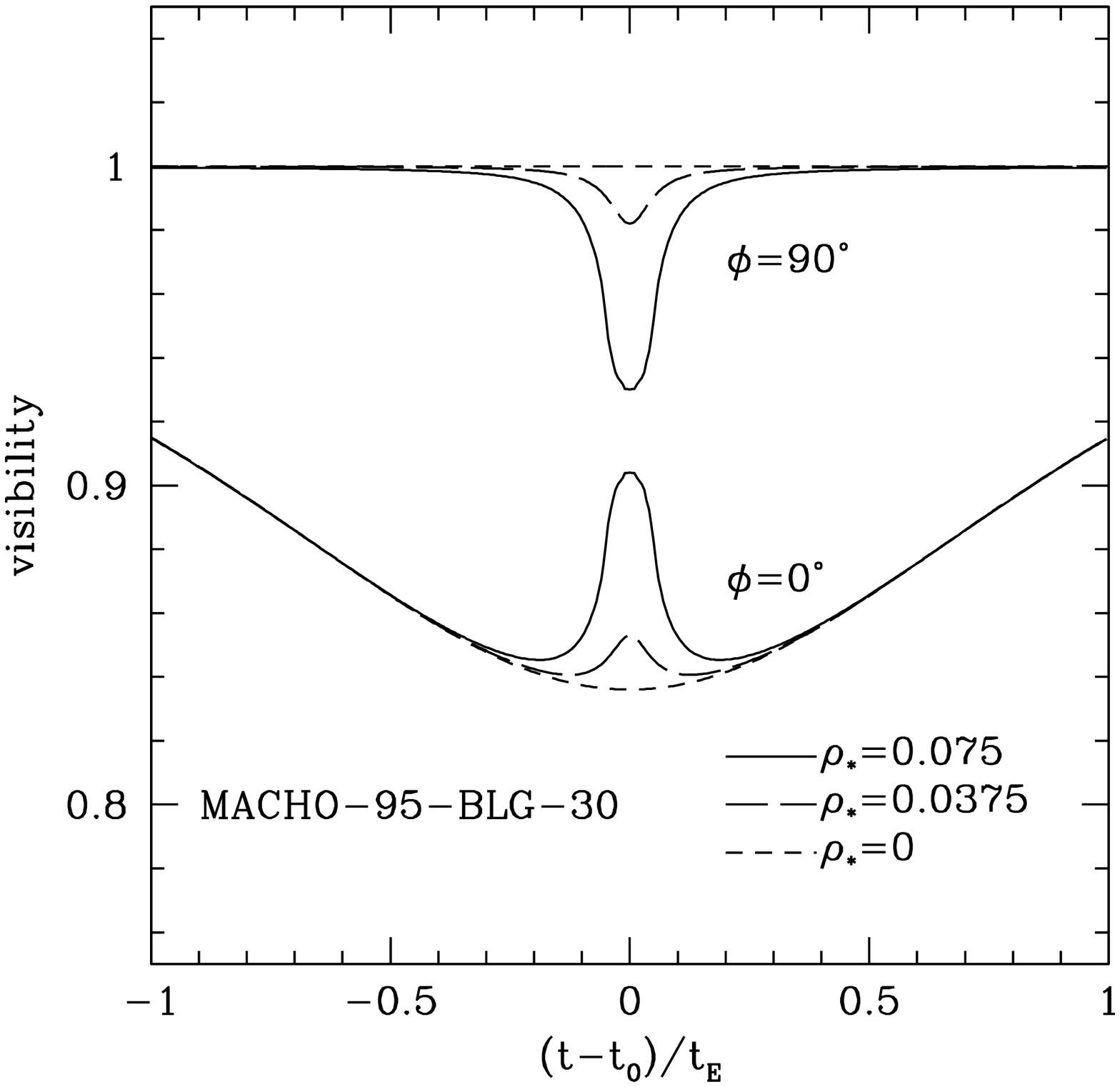}
\centering\includegraphics[width=0.49\hsize]{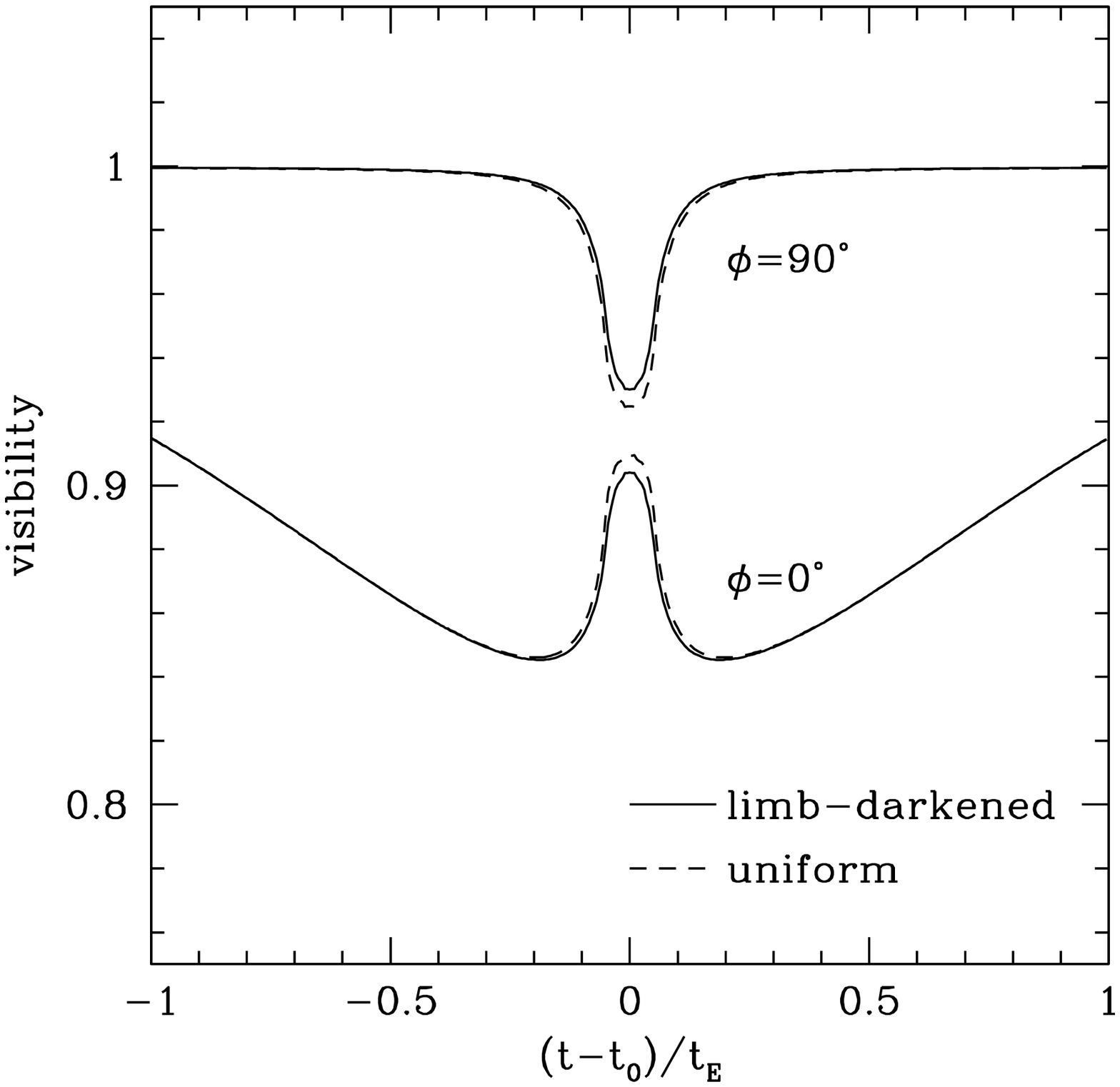}
\caption{
The left panel shows how visibility changes as a function of the source size
for which we take the limb-darkening profile of the MACHO-95-BLG-30
event in the MACHO R-band ($u_1=0.72, u_2=0.05$ as defined in eq. \ref{eq:limb}), (\citealt{Alc97}). 
The horizontal axis shows time in normalised units: $(t - t_{0})/t_{E}$,
where $t_{0}$ is the time of lensing maximum and $t_{E}$ is the
Einstein-radius crossing time (68 days in this example).
Three source sizes (in units of the Einstein radius) are shown, $\rs=0.075$, 0.0375 and 0 (point source).
The top and bottom curves are respectively for the cases where the baseline is
always perpendicular and parallel to the two micro-images.
The right panel shows how visibility changes as a function of the source
brightness profile. The solid lines assume a limb-darkening profile 
as before, 
while the dashed lines show the predictions for a uniform surface brightness profile.
The top and bottom curves are for the cases where the interferometer
baseline is always perpendicular to and parallel to the two micro-images.
}
\label{fig:test}
\end{figure}

The left panel of Fig. \ref{fig:test} shows how the visibility depends
on the source size for the event MACHO-95-BLG-30. For this event, the
minimum impact parameter is $\umin=0.05408$, and the critical
interferometer parameter is $\calK=0.68$. We assume three source sizes: 0.075 (as observed
for MACHO-95-BLG-30), 0.0375 and 0 (point source), and take the
limb-darkening parameters in the MACHO-R band (see Table \ref{table:single}).
For this exercise, we assume (rather artificially) that the
interferometer baseline can be continuously adjusted to be perpendicular or parallel
to the two micro-images. When the baseline is perpendicular to
the two micro-images, if the micro-images are modelled as points, the
visibility is always unity because there is no phase difference between
the two micro-images. However, when the source size is taken into account, the
visibility shows a decrement. For $\rs=0.075, 0.0375$, the decrement reaches
about 7\% and 2\%  respectively. As the VLTI is expected
to be sensitive to visibility changes of $\gtrsim$ 0.5\%, clearly, in this
case, the decrement caused by the finite source size effect makes the
event more observable.
For the case when the baseline is
parallel to the two micro-images, when the source size is zero, the
visibility decreases from unity to as low as 86\%. When the finite
source size is accounted for, the visibility shows a pronounced
deviation from the point-source approximation. The deviation lasts
for roughly two source diameter crossings in this case, and has an amplitude
of about 7\%, again readily observable.

The right panel of Fig. \ref{fig:test} shows
the change in visibility when we model the source with uniform
surface brightness rather than a limb-darkened profile. The interferometers
are again assumed to be adjusted to be either parallel or perpendicular
to the two micro-images.  In both cases, the change in the visibility
due to different surface brightness profiles is quite small, $\sim 1\%$, 
marginally detectable with the VLTI.

\begin{figure}
\centering\includegraphics[width=0.49\hsize]{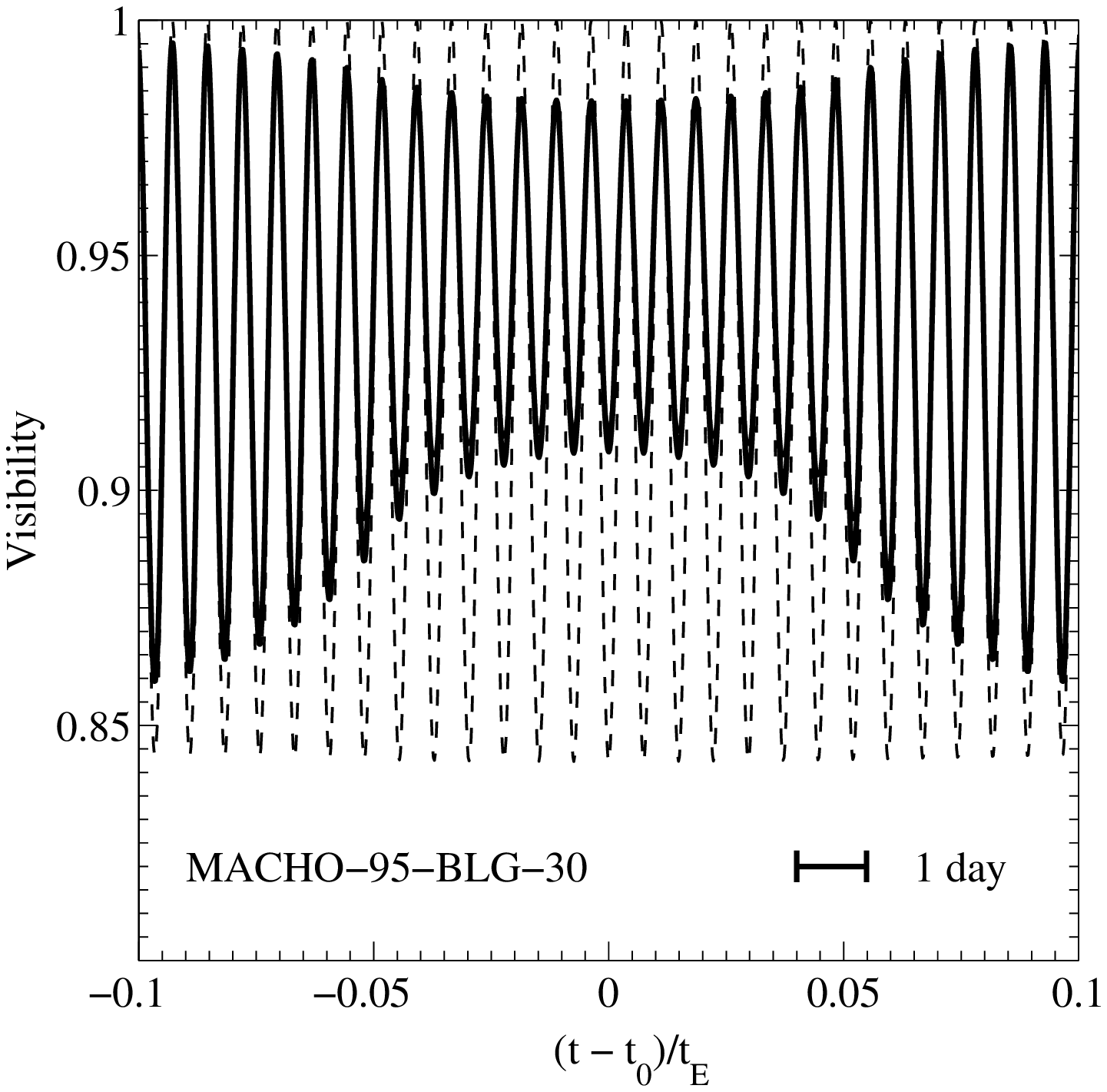}
\centering\includegraphics[width=0.49\hsize]{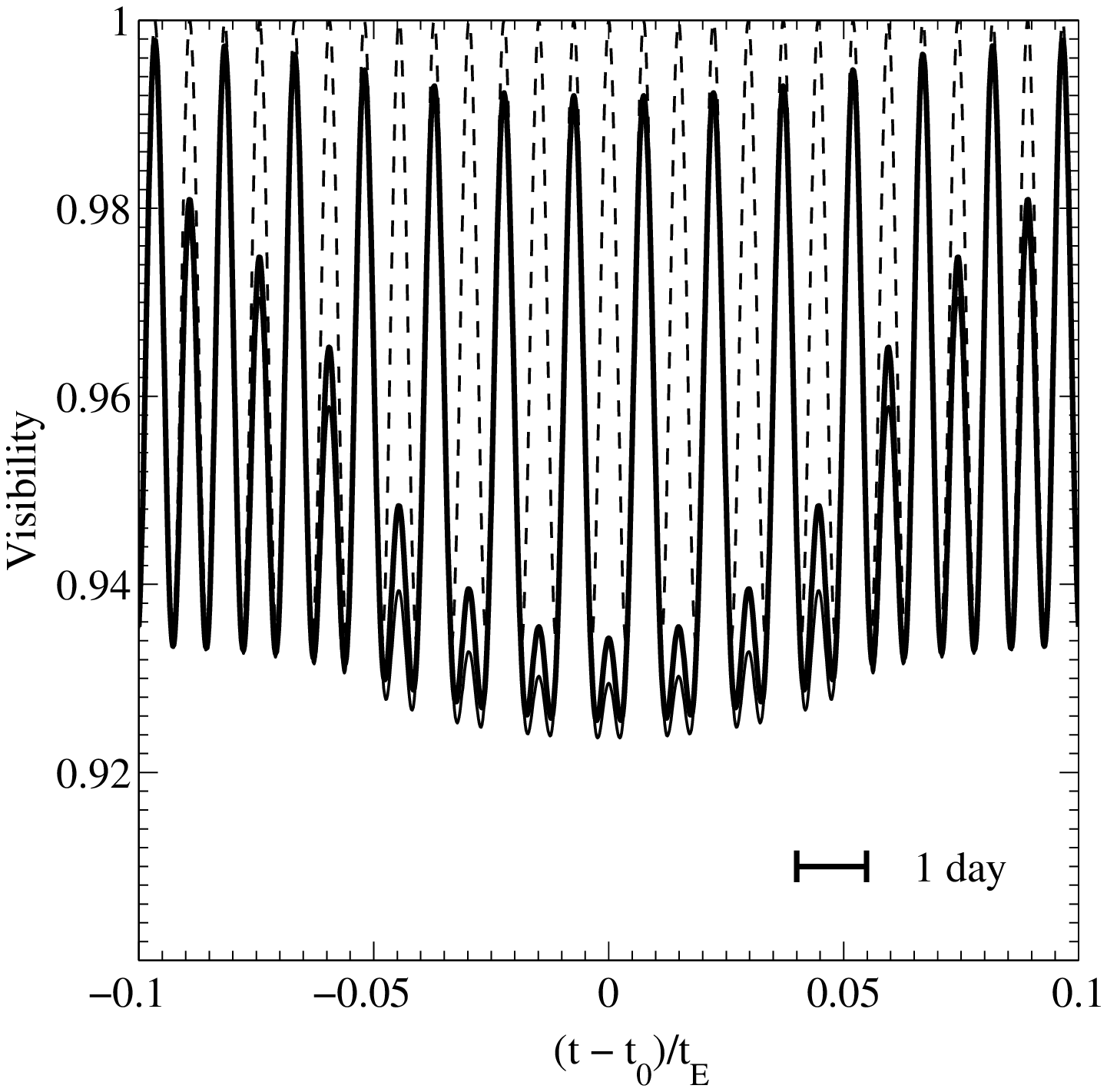}

\caption{
Theoretical visibility measurements for a finite source star with a 2 element interferometer. The separation between interferometer elements is 100m. The wavelength of observation is 2.2 $\mu$m. The event has a declination of $-30^{\circ}$ and the interferometer is at latitude $-40^{\circ}$. The hour angle at maximum lensing amplification is $0.0^{h}$. The event has microlensing parameters similar to the MACHO-95-BLG-30 event, with $u_{\rm min}=0.05408$ and $\rs=0.075$. The left and right panels show the visibility with a baseline aligned in the east-west and north-south directions respectively. The visibility for a  limb-darkened source and a uniform brightness source is shown by thick  and thin lines respectively. The visibility for a point-source is shown by the dashed line.
}
\label{fig:real_BL_9530}
\end{figure}

\begin{figure}
\centering\includegraphics[width=0.48\hsize]{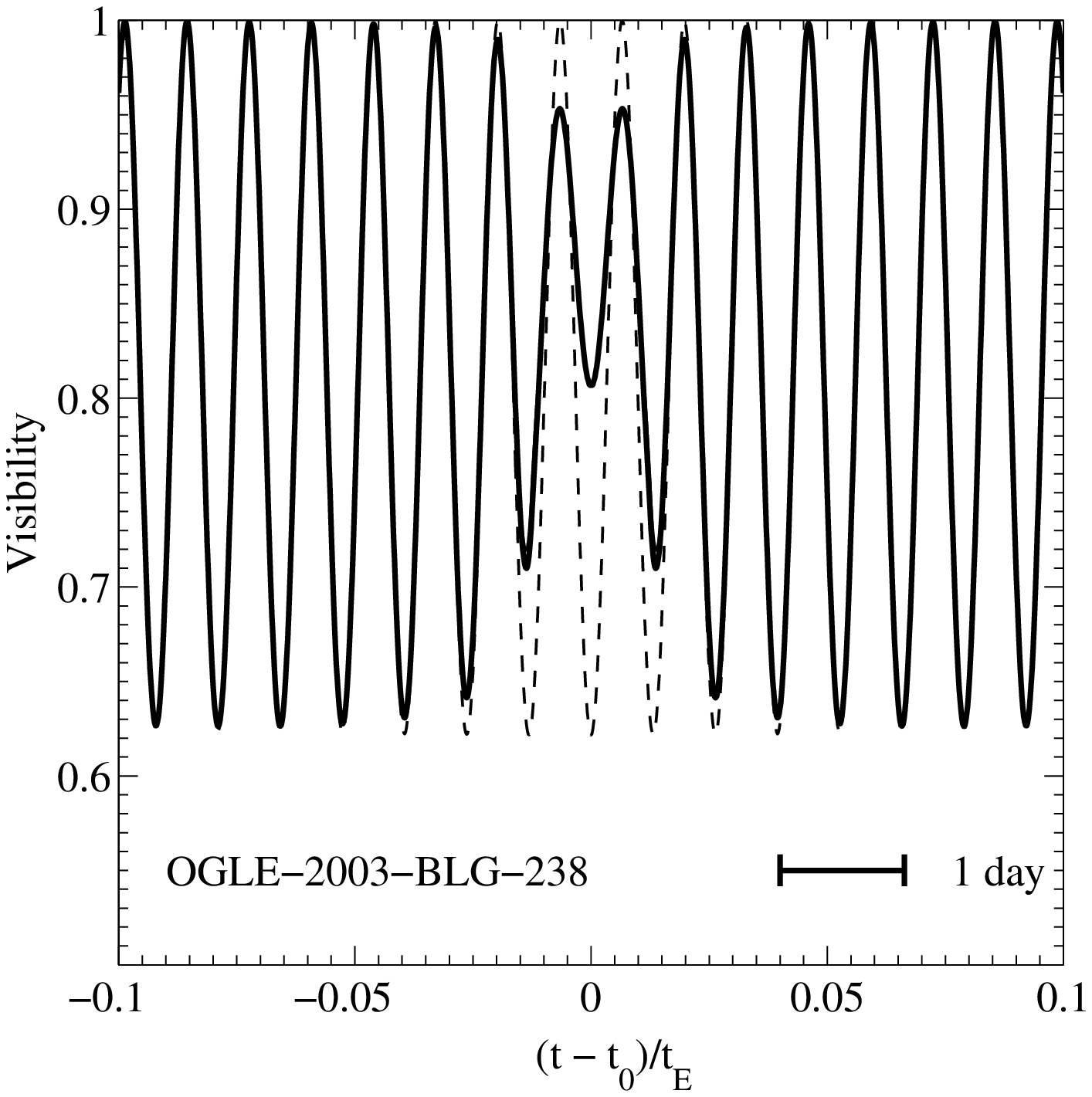}
\centering\includegraphics[width=0.49\hsize]{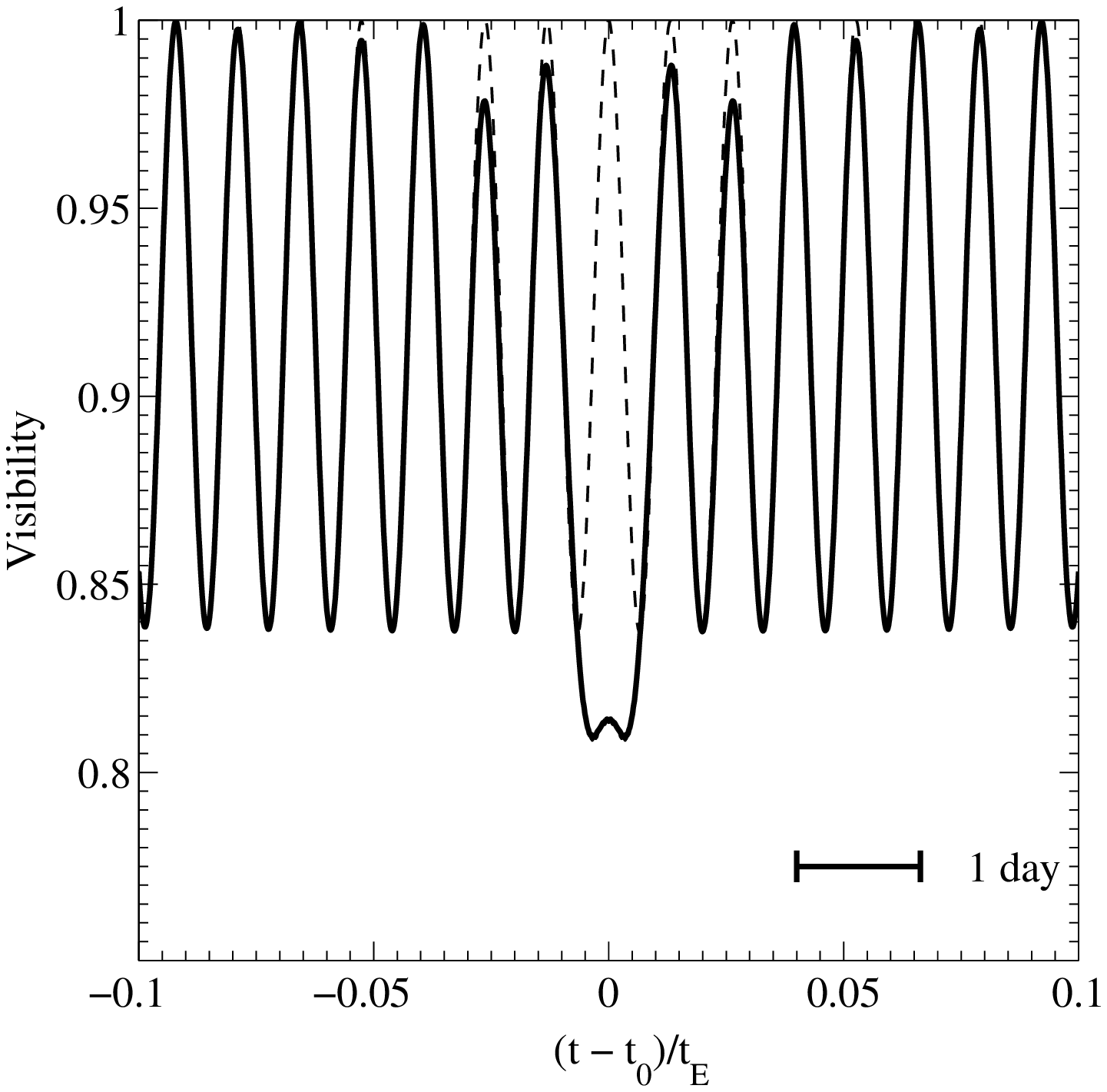}

\caption{
Theoretical visibility measurements for event OGLE-2003-BLG-238. The interferometer parameters are as for Fig.~\ref{fig:real_BL_9530}, and the event parameters are listed in Table~\ref{table:single}.
}
\label{fig:real_BL_238}
\end{figure}

\begin{figure}
\centering\includegraphics[width=0.48\hsize]{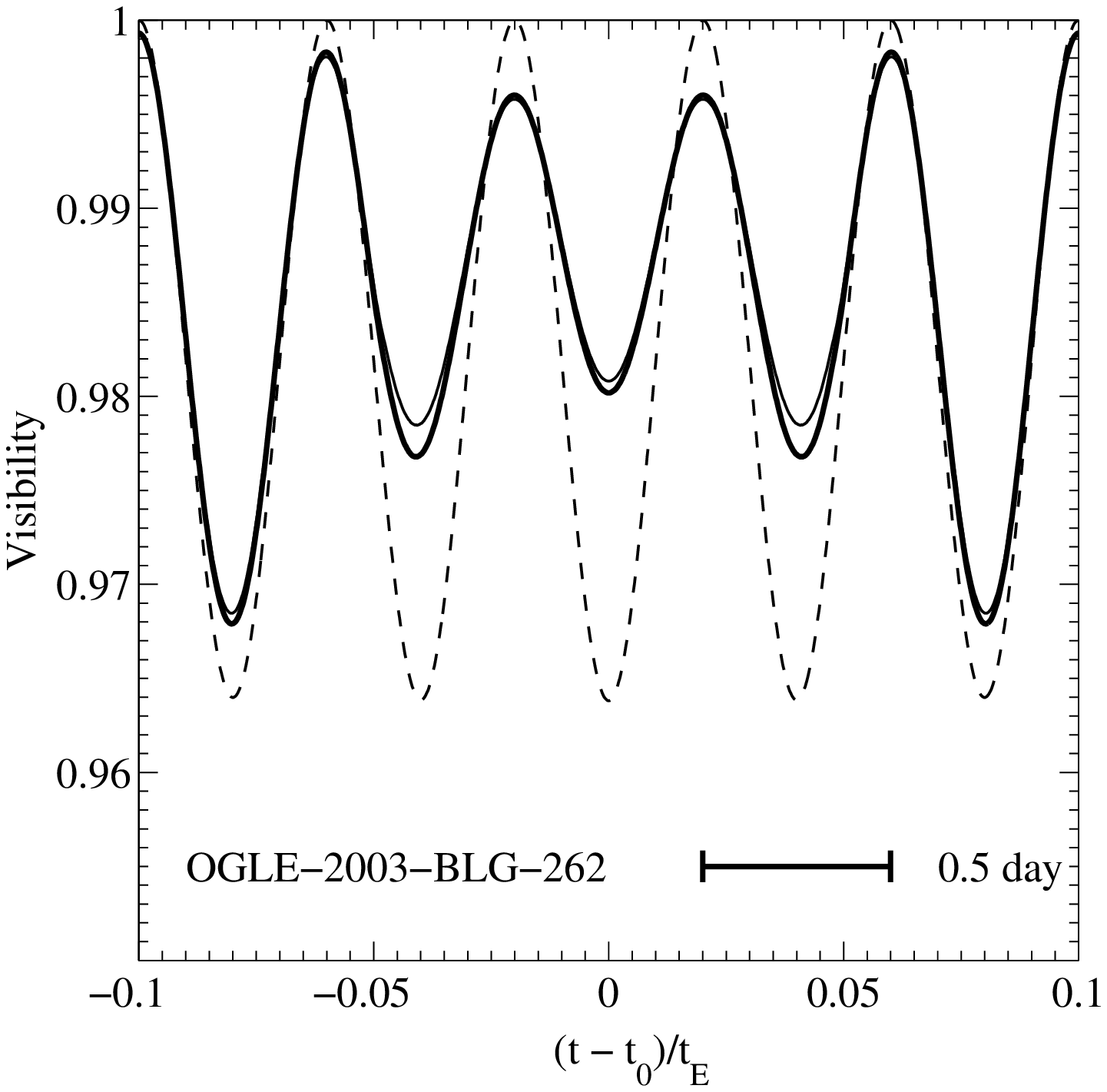}
\centering\includegraphics[width=0.49\hsize]{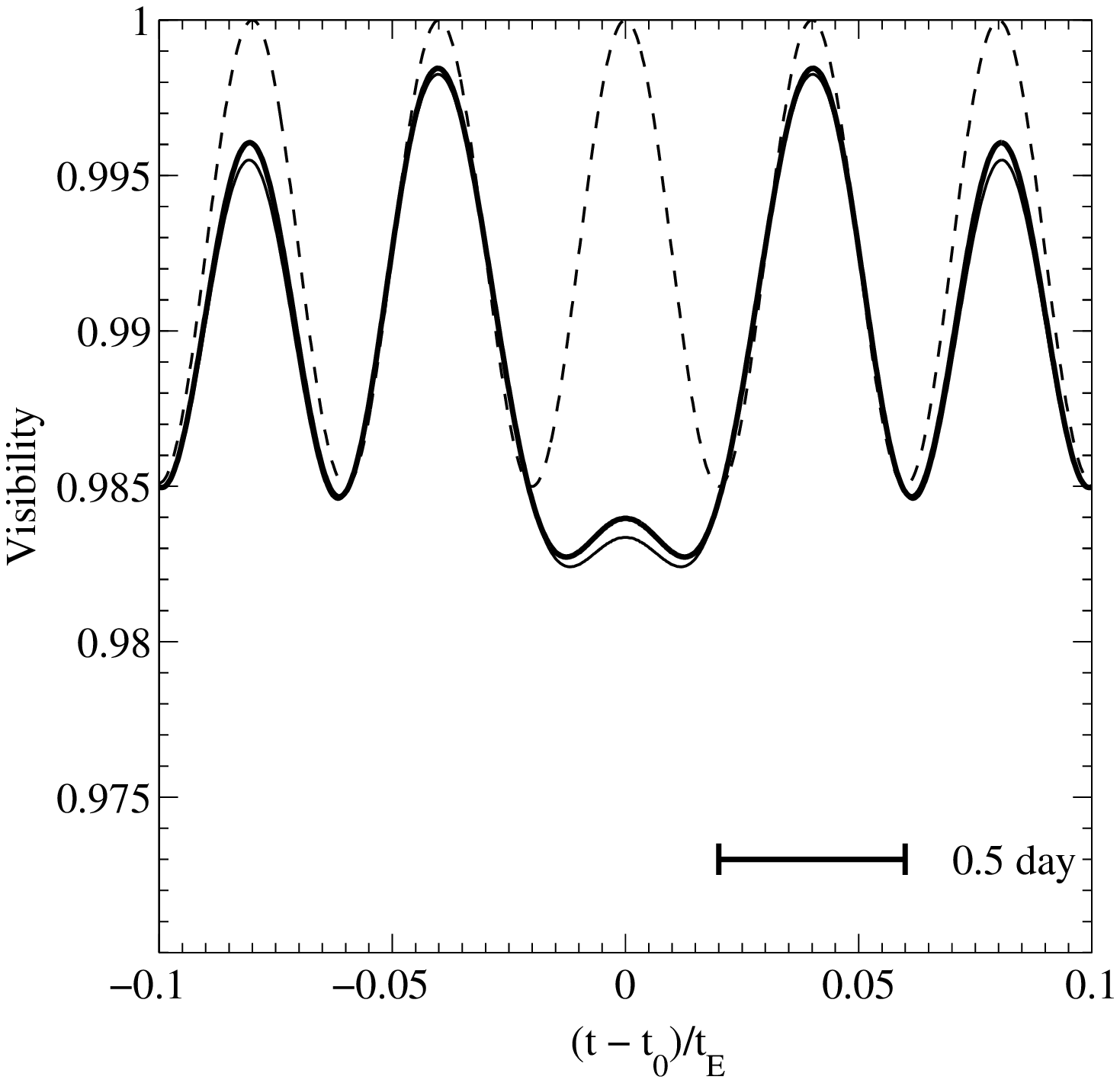}

\caption{
Theoretical visibility measurements for event OGLE-2003-BLG-262. The interferometer parameters are as for Fig.~\ref{fig:real_BL_9530}, and the event parameters are listed in Table~\ref{table:single}.
}
\label{fig:real_BL_262}
\end{figure}

In reality, we do not know how the two micro-images are aligned relative
to the interferometers. We now adopt a more realistic simulation where we allow the baseline to move according to the Earth's rotation. Figures~\ref{fig:real_BL_9530}, \ref{fig:real_BL_238} and \ref{fig:real_BL_262} show the visibilities for the three microlensing events described in Table~\ref{table:single}. The effect of Earth rotation is immediately obvious as the visibility undergoes cyclic changes. In each figure the visibilities are shown for baselines oriented in the east-west and north-south directions. In these simulations the E-W (N-S) baseline is parallel (perpendicular) to the image separation at the event maximum. The visibilities for uniformly bright and limb-darkened source stars are shown in comparison to those assuming a point source. It is clear that for any of the three microlensing events, the visibilities obtained by two elements of the VLTI are significantly different to those expected from a point source, at times near the event maximum. At times prior to, and after peak magnification, the visibility curves tend toward the point-source results. This is as expected as the distortion of the source images is greater at higher magnification.  The difference in visibility due to a finite source  for each of the events exceeds the minimum difference detectable by the VLTI for a time roughly equal to twice the source diameter crossing time.

\subsection{Three-element interferometer}
\begin{figure}
\centering\includegraphics[width=0.49\hsize]{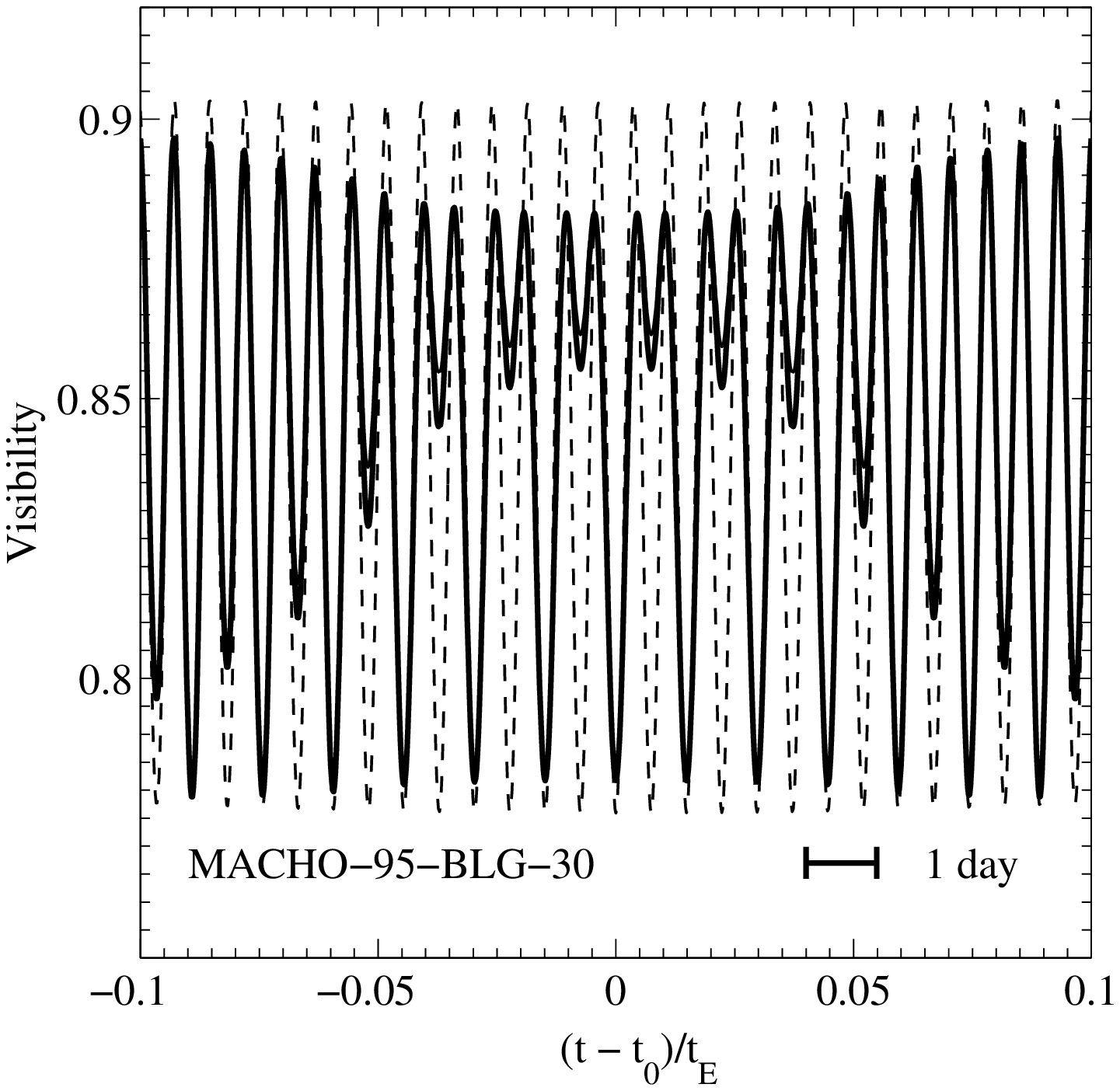}
\centering\includegraphics[width=0.4875\hsize]{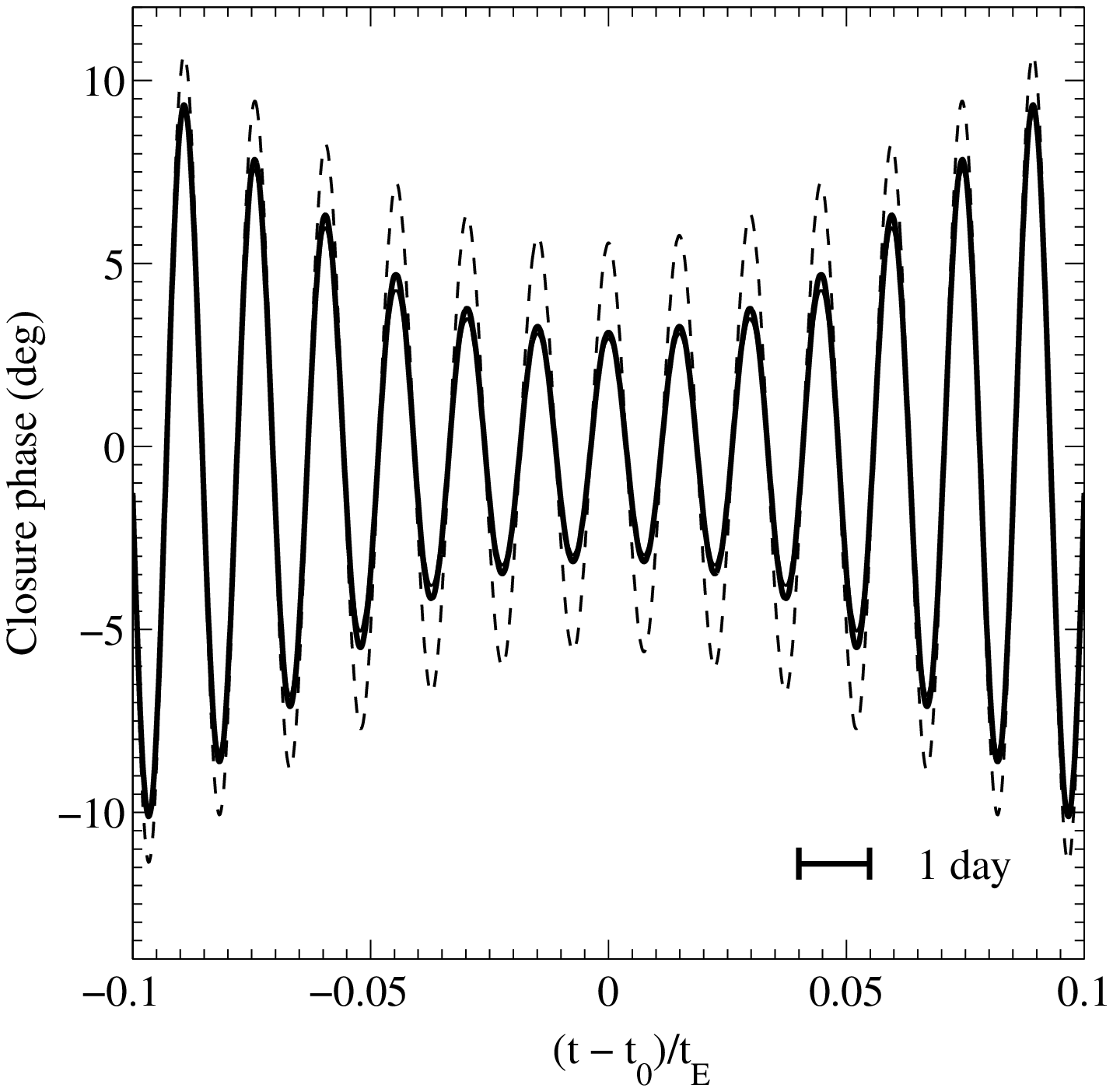}

\caption{
Theoretical measurements using a three-element interferometer for event MACHO-95-BLG-30. The interferometer baselines are arranged as an equilateral triangle with side length $B = 100$m. The other interferometer parameters are as given for Fig.~\ref{fig:real_BL_9530}. The microlensing event parameters are listed in Table~\ref{table:single}. Left: The product $|V_{1}|\,|V_{2}|\,|V_{3}|$, see text. Right: The closure phase $(\phi_{12} + \phi_{23} + \phi_{31})$. The results are shown for a limb-darkened source (thick line), uniformly bright source (thin line) and a point source (dashed line).
}
\label{fig:3EL_9530}
\end{figure}

\begin{figure}
\centering\includegraphics[width=0.485\hsize]{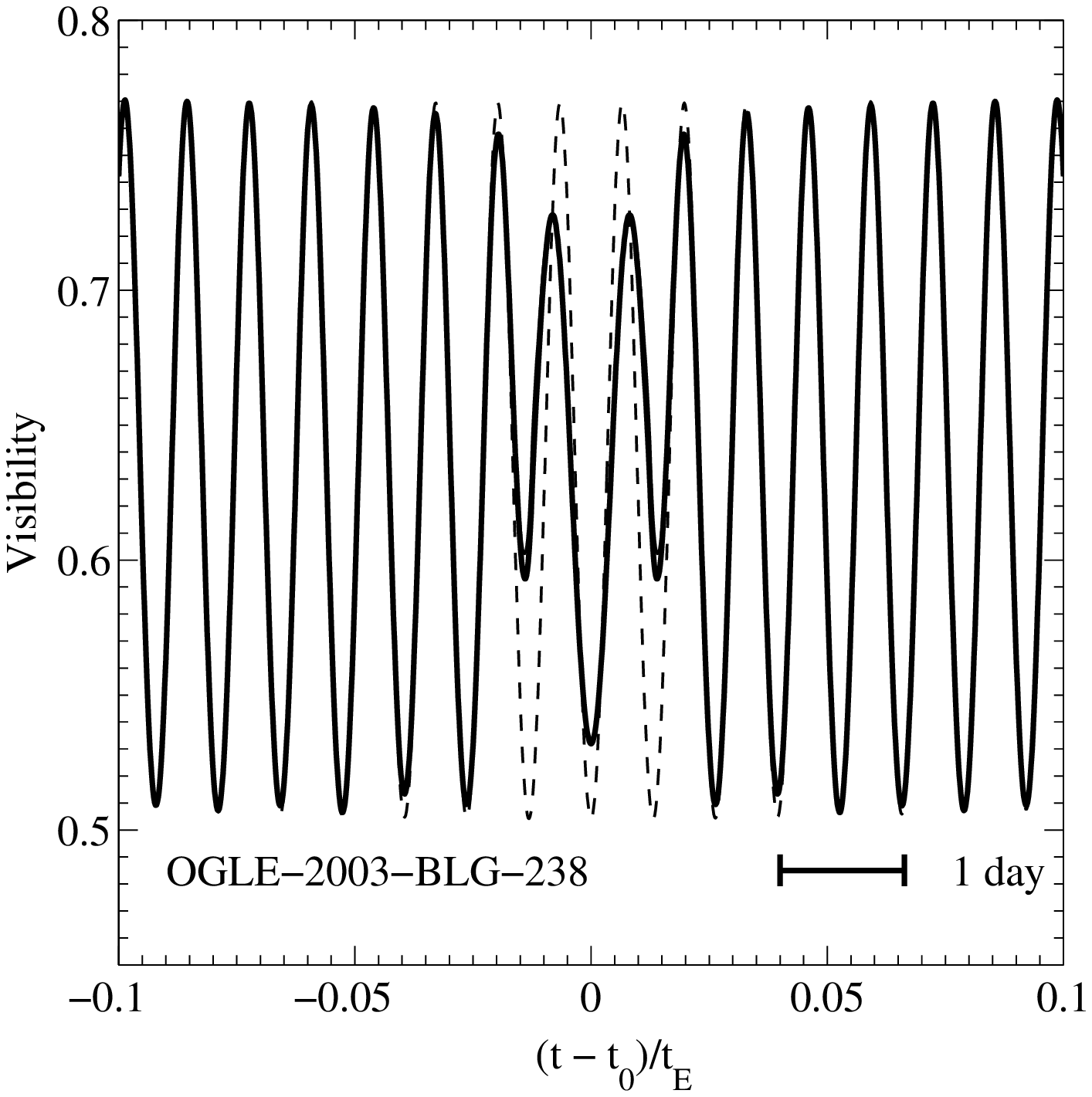}
\centering\includegraphics[width=0.492\hsize]{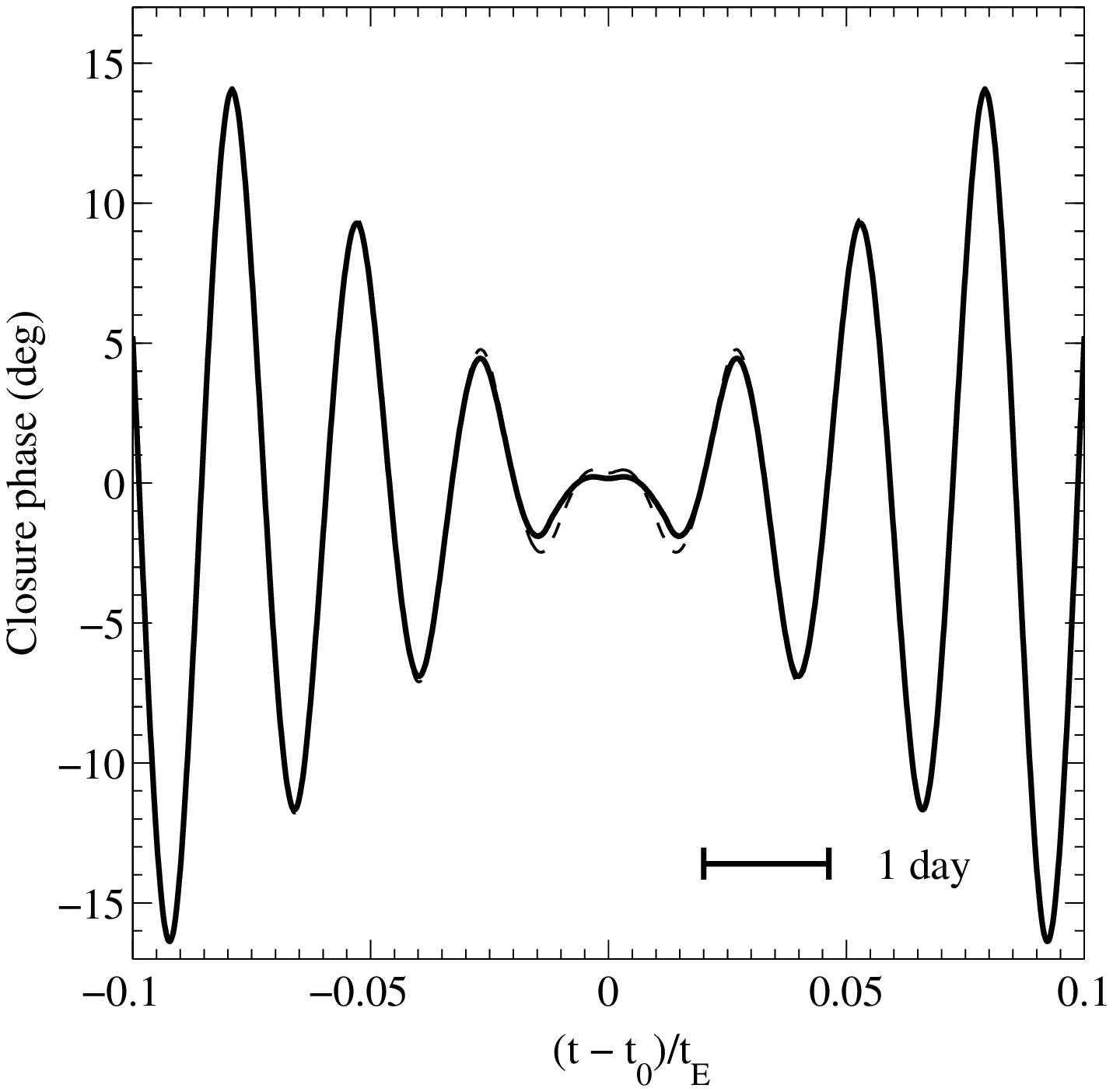}

\caption{
Theoretical measurements using a three-element interferometer for event OGLE-2003-BLG-238. Left: The product $|V_{1}|\,|V_{2}|\,|V_{3}|$, see text. Right: The closure phase. The results are shown for a limb-darkened source (thick line), uniformly bright source (thin line) and a point source (dashed line).
}
\label{fig:3EL_238}
\end{figure}

\begin{figure}
\centering\includegraphics[width=0.4925\hsize]{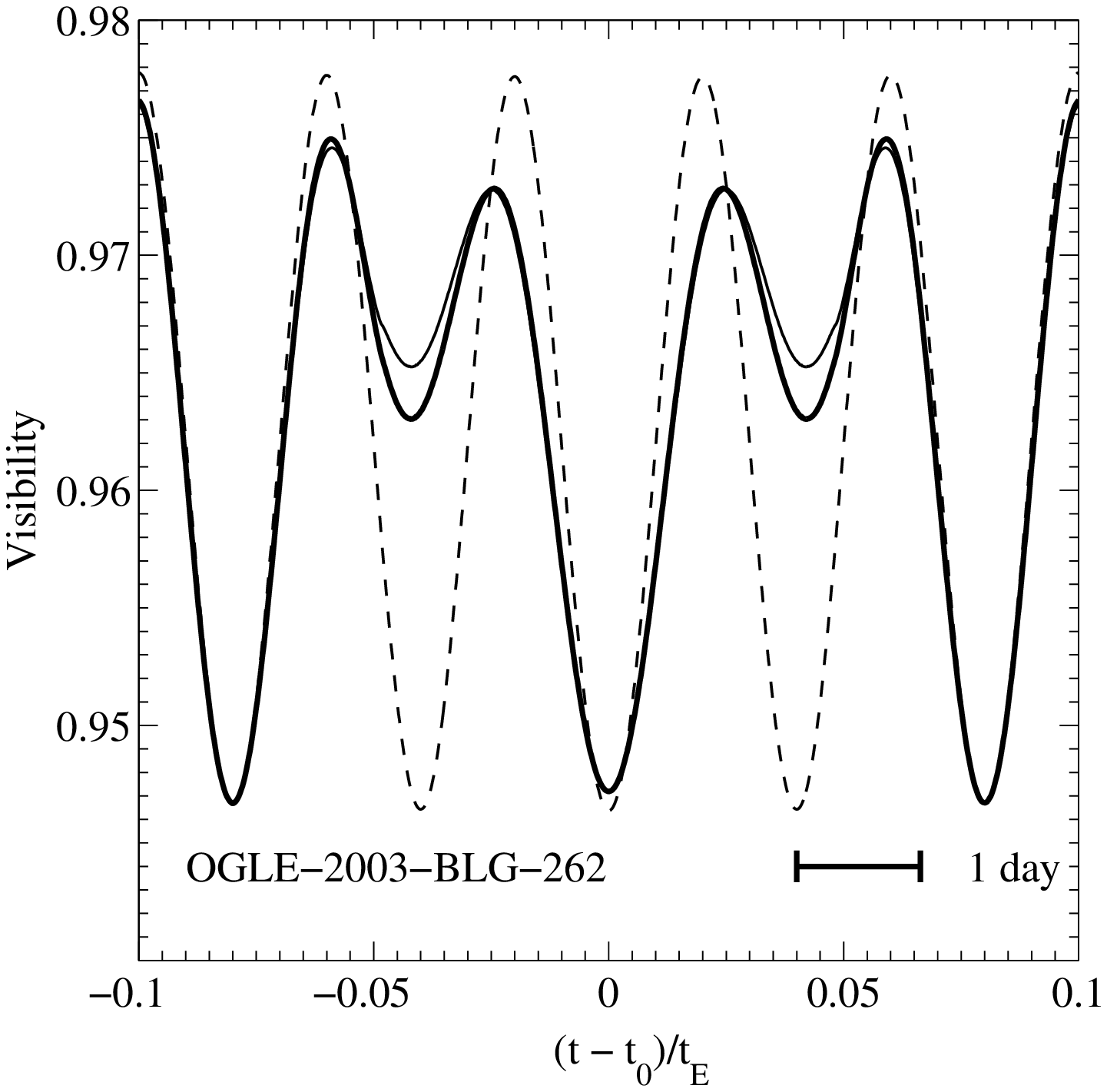}
\centering\includegraphics[width=0.48\hsize]{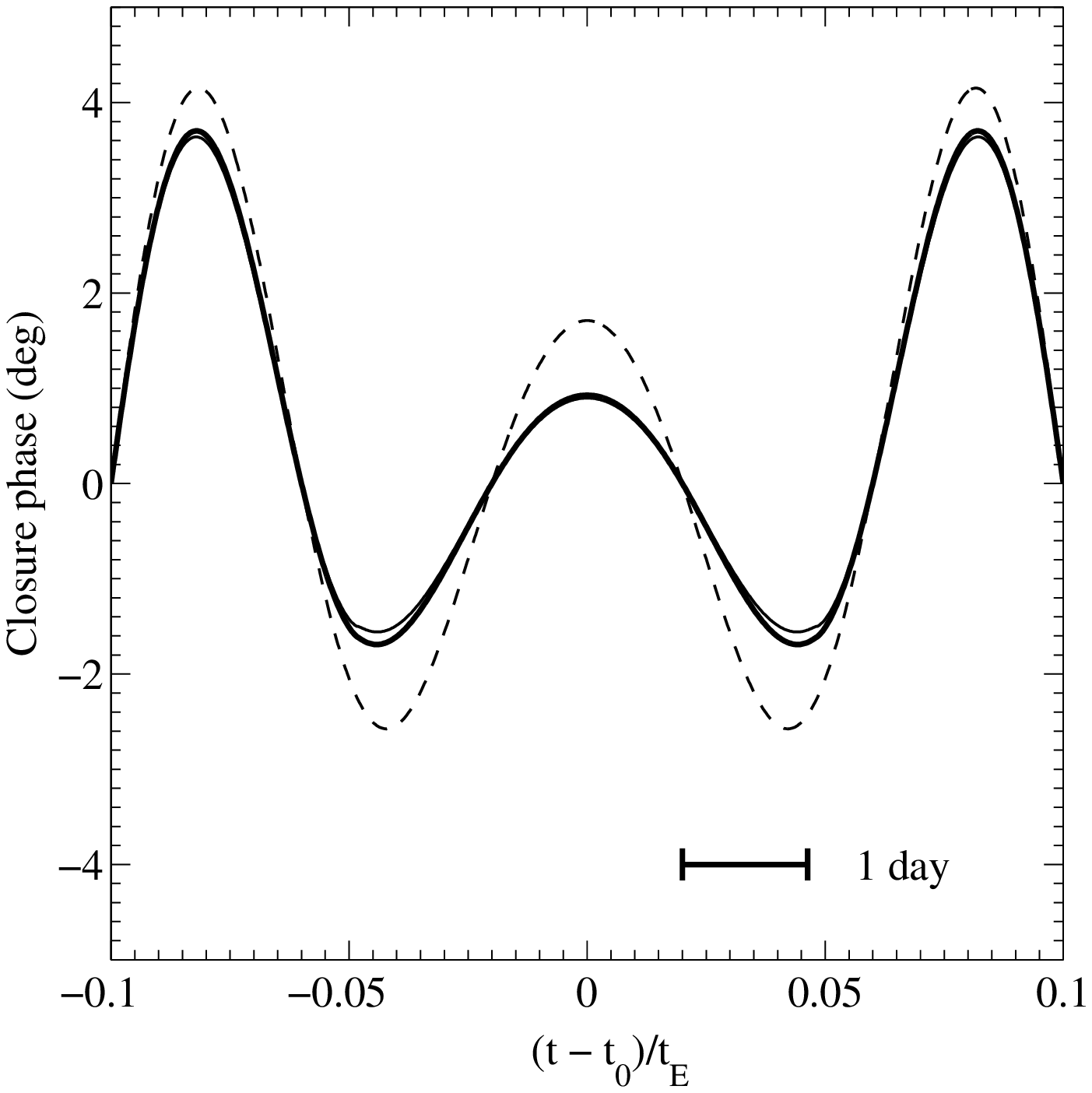}

\caption{
Theoretical measurements using a three-element interferometer for event OGLE-2003-BLG-262. Left: The product $|V_{1}|\,|V_{2}|\,|V_{3}|$, see text. Right: The closure phase. The results are shown for a limb-darkened source (thick line), uniformly bright source (thin line) and a point source (dashed line).
}
\label{fig:3EL_262}
\end{figure}

We perform the simulations again, using a three-element interferometer with the baselines arranged as an equilateral triangle with baseline length 100m. All other telescope parameters are as above. Figures~\ref{fig:3EL_9530}, \ref{fig:3EL_238} and \ref{fig:3EL_262} show the product of the visibilities from each baseline, $|V_{1}||V_{2}||V_{3}|$, as well as the closure phase for each of the three events listed in Table~\ref{table:single}. The source is modelled with limb-darkened and uniform brightness profiles as before. These are compared to the results assuming a point-source. The effect of a finite source star is again evident in the visibility curves around the time of maximum magnification for a time roughly equal to twice the source diameter crossing time. The difference in closure phase due to a finite source exceeds the minimum value detectable by VLTI ($1^{\circ}$) for all events other than OGLE-2003-BLG-238.

\begin{figure}
\psfrag{xlabel}{\large  \mbox{$\log_{10} (\rs)$}  }
\centering\includegraphics[width=0.505\hsize]{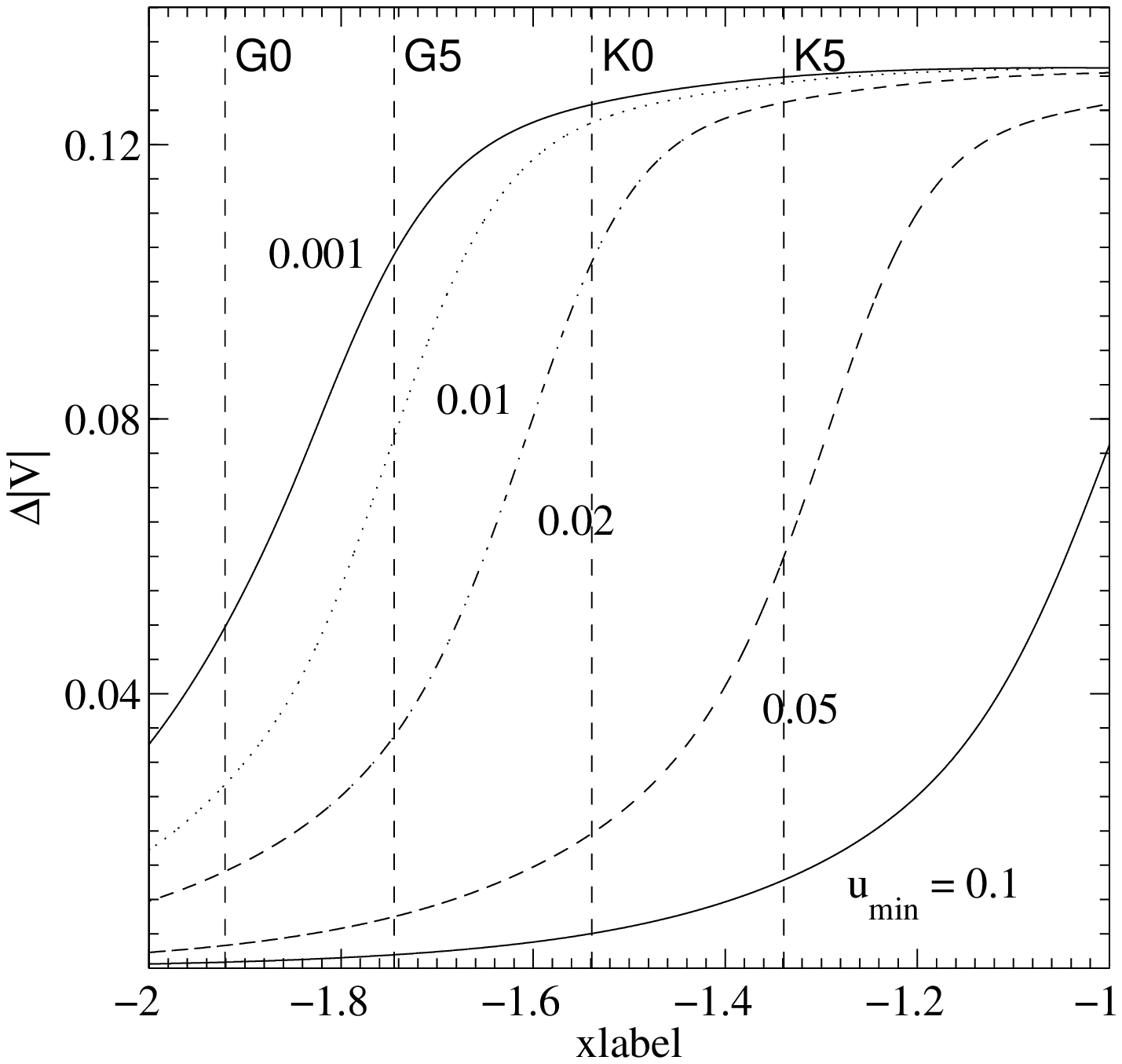}
\centering\includegraphics[width=0.485\hsize]{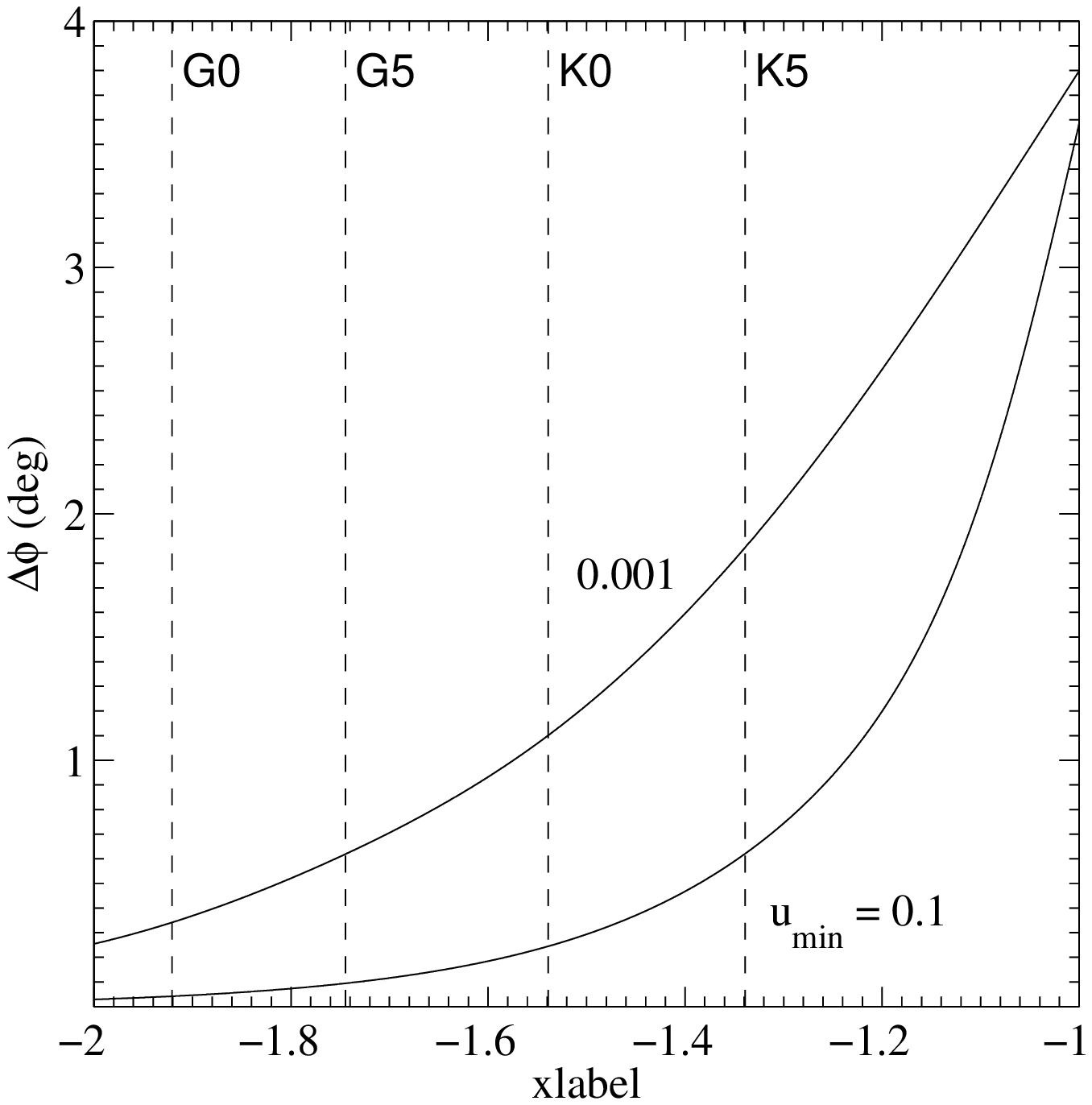}

\caption{\label{fig:estimates}The maximum effect of finite source size on visibility and closure phase as a function of  source star size, for various values of minimum impact parameter, $\umin$. The parameters of the three-element interferometer are described in the caption of Fig.~\ref{fig:3EL_9530}. Left: The fractional change in visibility as a function of the normalised stellar radius, $\rs \equiv {\thetas / \thetaE}$. The top and bottom solid lines correspond to microlensing events with  $\umin = 0.001$ and 0.1 respectively. Other values of $\umin$ were used: 0.05 (dashed), 0.02 (dot-dashed), 0.01 (dotted). Right: The absolute change in closure phase owing to finite source effects. The top and bottom lines correspond to $\umin = 0.001$ and 0.1 respectively. The approximate source star sizes for K and G type giants at 8 kpc are shown as vertical lines, assuming a 0.3 solar mass lens star at 6 kpc.}
\end{figure}

The fraction of microlensing events that will be affected by finite source effects may reach $\sim 3\%$ (\citealt{Wit95}, see also \citealt{Gou94}).  The parameters which govern how the visibility and closure phase differs for a finite source with respect to a point source are $\rs$ and $\umin$. Figure~\ref{fig:estimates} shows the maximium difference in these observables using a finite source star compared to a point source, as a function of $\rs$ and $\umin$. We assume a lens mass of $0.3M_{\odot}$ at a distance 6 kpc, with the source star at 8 kpc. We consider a range of $\rs$ corresponding to K and G type giant stars. We note that the maximum difference in visibility due to a finite source star is approximately 13\%. The maximum difference in closure phase is approximately 4 degrees. These should be compared with the expected accuracy of the VLTI of $\sim 0.5 - 5\%$ and $\sim 1^{\circ}$ respectively.

\section{Summary and Discussion}

In this paper, we have studied how the interferometric fringe visibility
of a microlensing event is affected by the finite size of the source star.
When we take into account realistic telescope sites and
interferometer baselines, the visibility undergoes sinusoidal changes, as illustrated in Figures~\ref{fig:real_BL_9530}--\ref{fig:3EL_262}.
Therefore it will be useful to observe the fringe
visibilities during multiple epochs for a microlensing event.

We have shown that the fringe visibility can change by as much as $\sim$ 10\%, which
can be easily measured by the VLTI.  Unfortunately, it
will be challenging to study limb-darkening profile parameters using interferometers.
 The ratio $\rs = \theta_{\rm s} / \theta_{\rm E}$ can be accurately determined via the modelling of the photometric microlensing light curve.
Combined with the
source colour, this allows us an approximate determination of the source size, $\theta_{\rm s}$. and thereby the Einstein ring radius $\theta_{\rm E}$.
The interferometric finite source size, on the other hand, determines
both the angular Einstein radius and the finite source size directly, without
reference to the source colour. So the effect of a finite source star size on the fringe visibility 
complements the photometric finite source size effect. Observations
of finite source size events may also be important for another reason: 
for these events, we already have some indication of the source size,
so they can be used to verify the interferometer capabilities, which may 
be particularly important at early stages.

In addition, the visibility measurement gives us the source trajectory relative to
the interferometer baseline. This additional information can be used
to further constrain the lens and source kinematics in a maximum
likelihood analysis of the lens and source locations.

For a two-element interferometer, when
the baseline is perpendicular to the two images, the visibility is
always unity so the two micro-images cannot be resolved.
This is no longer the case when we have more
than two elements, as there are always a baseline that is at some angle to
the two micro-images, so the visibility decrement can be
observed. Therefore a multiple-element interferometer has a
substantial advantage over a two-element interferometer.

The finite source size effect lasts roughly a factor of two 
times the source diameter crossing time.
For the three events listed in Table \ref{table:single}, the source
diameter crossing times are  about 10, 1 and 1.5 days for  MACHO-95-BLG-30,
OGLE-2003-BLG-238, and OGLE-2003-BLG-262 respectively, so
there should be sufficient time to use interferometers
to gain further insights on the lensing parameters. As the finite source
size effects can both enhance and decrease the visibility and closure
phases, it will be helpful to perform at least two observations: one
around the peak of light curve and one shortly afterwards when the
finite source size effect is no longer important. 

In this paper, we have not studied binary lensing events under either
a point source approximation or with realistic finite source sizes.
Some of these events will have
large magnifications during a caustic crossing. It will be interesting
to study the interferometric signals of this class of events, and how
the visibility changes during a binary (including planetary) lensing
event, and what extra information one can extract.

We thank Ian Browne and Neal Jackson for many helpful discussions
about interferometry; Bohdan Paczy\'{n}ski, Duncan Lorimer, Andrea Richichi, Phil Yock and the referee for their comments on this manuscript. NJR is supported a PPARC postdoctoral grant. 
This work was partially supported by the European Community's Sixth
Framework Marie Curie Research Training Network Programme, Contract
No. MRTN-CT-2004-505183 `ANGLES'.

\appendix

\section{Fringe Visibility when the source is exactly along the line of sight}

When the source is exactly aligned with the line of sight, the images form an annulus.
The inner and outer radii in units of the Einstein radius are given by
\beq
r_{+, -}={\sqrt{\rs^2+4} \pm \rs \over 2},
\eeq
where $\rs$ is the physical source size in units of the Einstein radius.

To derive the visibility, it is natural to use the cylindrical
coordinate system due to symmetry.
Without losing generality, we can put the interferometer baseline along
the  horizontal axis. Let us assume the source has uniform surface brightness. Recall that
$\vec{\cal K}_{\rm cr} \cdot \vec{r} = \calK r \cos\theta$, where $\theta$ is the polar
angle, and $\calK=2 \pi B \thetaE/\lambda$. The complex visibility is
given by
\beq
\hat{V} ={1 \over \int_{r_-}^{r_+} 2 \pi r dr} \int_{r_-}^{r_+} r dr \int_0^{2 \pi} \exp(i \calK r \cos\theta) d\theta
= 2\,{ r_+ J_1(\calK r_+) - r_- J_1(\calK r_-) \over \calK (r_+^2 - r_-^2)},
\eeq
where $J_n(x)$ is the usual Bessel function of order $n$. Notice that due
to symmetry, the visibility $\hat{V}(\calK)$ is real. 

For an infinitesimal source, i.e., at the limit $\rs\rightarrow 0$, we find
\beq
\hat{V} = {J_1(\calK) \over {\calK}} +{J_0(\calK)-J_2(\calK) \over 2}.
\eeq

\bibliographystyle{mn2e}

\end{document}